\documentclass[prb,twocolumn,showpacs,amsmath,amssymb]{revtex4}

\usepackage{graphicx}
\usepackage{dcolumn}
\usepackage{bm}
\usepackage{epsfig}

\begin{document}

\title{Mesoscopic Spin Hall Effect in Multiprobe Ballistic Spin-Orbit 
Coupled  Semiconductor Bridges}

\author{Branislav K. Nikoli\' c, Liviu P. Z\^ arbo, and Satofumi Souma}
\affiliation{Department of Physics and Astronomy, University
of Delaware, Newark, DE 19716-2570, USA}

\begin{abstract}
We predict that unpolarized charge current driven through the longitudinal leads 
attached to ballistic  quantum-coherent  two-dimensional electron gas (2DEG) 
in semiconductor heterostructure will induce a {\em pure} spin current, which is not accompanied 
by any net charge  flow, in the transverse voltage probes. Its magnitude can be tuned by the Rashba 
spin-orbit (SO)  interaction and, moreover, it is resilient to weak spin-independent scattering off 
impurities  within the metallic  diffusive regime. While the polarization vector of the spin transported through the transverse leads is not orthogonal to the plane of 2DEG, we demonstrate that only two 
components (out-of-plane and longitudinal) of the transverse spin current  are
signatures of the spin Hall effect in four-probe Rashba spin-split semiconductor nanostructures. The linear response spin Hall current, obtained from the multiprobe Landauer-B\" uttiker scattering formalism generalized for quantum transport of spin, is the Fermi-surface determined nonequilibrium quantity whose scaling with 
the 2DEG size $L$ reveals the importance of processes occurring  on the spin precession {\em mesoscale} $L_{\rm SO}$ (on which spin precesses by an angle $\pi$)---the out-of-plane component of 
the transverse spin current exhibits quasioscillatory behavior for $L \lesssim L_{\rm SO}$ (attaining  
the  maximum value in 2DEGs of the  size $L_{\rm SO} \times L_{\rm SO}$), while it reaches the 
asymptotic value in the macroscopic regime $L \gg L_{\rm SO}$. Furthermore, these values of the spin Hall 
current can be manipulated by the measuring geometry defined by  the attached leads.
\end{abstract}

\maketitle

\section{Introduction}

Current efforts in spintronics are to a large extent 
directed toward gaining control of electron spin in  semiconductor 
structures, which are ubiquitous in conventional electronics, and exploiting 
it as a carrier of classical or quantum information.~\cite{spintronics} 
Although spin physics in semiconductors is an old subject,~\cite{rashba,extrinsic_1}  
spintronics has reignited interest in the role of spin-orbit (SO) couplings   
in condensed matter systems.  While they originate from relativistic corrections to 
the Schr\" odinger equation, SO interactions  for itinerant  electrons in semiconductor 
nanostructures can be much stronger than for particles  moving through electric fields 
in vacuum.~\cite{rashba_review,winkler} Therefore, they are envisaged to be a tool 
for all-electrical~\cite{rashba_review,nitta} spin current generation and manipulation, where 
electric fields can be produced to control electron spin in far smaller volumes and on far shorter 
time scales than it is possible with conventional magnetic field-based spin control.~\cite{spintronics}

\begin{figure}
\centerline{\psfig{file=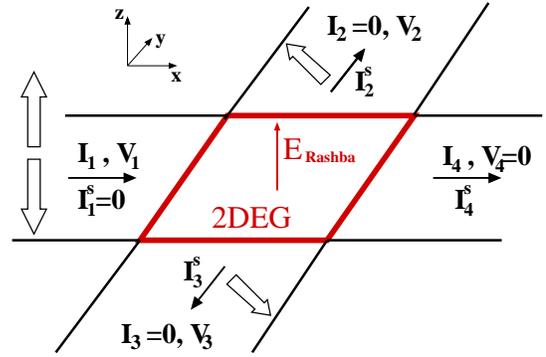,width=2.8in,angle=0}}
\caption{(Color online) The four-probe mesoscopic bridge for the detection of
pure spin Hall currents. The central region is 2DEG  where
electrons are confined within a semiconductor heterostructure by an
electric field  along the $z$-axis which  induces the Rashba SO
coupling. The four attached leads are clean, non-magnetic, and without any 
SO coupling. The unpolarized ($I_1^s=0$) charge current ($I_1 \neq 0$)  through the 
longitudinal leads induces spin Hall current in the transverse
leads which act as the voltage probes $V_2=V_3 \neq 0$, $I_2=I_3=0$.
We predict that the polarization vector of the spin transported by pure 
($I_2=I_2^\uparrow+I_2^\downarrow=0)$ spin current $I_2^s=\frac{\hbar}{2e}(I_2^\uparrow-I_2^\downarrow)$
is not orthogonal to the plane of 2DEG.} \label{fig:setup}
\end{figure}

Recent theoretically unearthed {\em intrinsic} spin Hall effect in hole doped~\cite{murakami} (such as bulk $p$-GaAs or $p$-Ge described by the Luttinger effective  Hamiltonian for heavy and light holes), or electron doped~\cite{sinova} (such as 2DEG in $n$-type heterostructures with structure inversion asymmetry which gives rise to the Rashba-type of SO coupling~\cite{rashba,rashba_review,winkler}) semiconductors  suggests that {\em pure} (i.e., not accompanied by any net dissipative charge current)  transverse spin current could be generated in these systems due to longitudinal electric field. The correlation between spin orientation and carrier velocity, induced in this effect in the presence of an external electric field, essentially requires some type of SO coupling which is strong enough to spin-split the Bloch energy bands.~\cite{murakami}  

On the other hand, it has been known for a long time~\cite{extrinsic_1,extrinsic_2} that SO dependent scattering off impurities will deflect spin-$\uparrow$ (spin-$\downarrow$) electrons predominantly to 
the right  (left), thereby giving rise to the {\em extrinsic} spin Hall effect where pure spin current flows  perpendicular to the longitudinal unpolarized charge transport. However, the intrinsic spin Hall current is 
expected to be several order of magnitude larger than the extrinsic one, thereby promising all-electrical solution to spin injection problems~\cite{rashba_review} and opening new avenues for semiconductor spintronics applications.~\cite{spintronics} 

While the properties of the intrinsic spin Hall current have been delineated through  
the semiclassical analysis of infinite homogeneous SO coupled semiconductor systems 
in the clean limit,~\cite{sinova,murakami}  guiding experimental detection of such 
effects requires a quantitative prediction for the spin current flowing through the 
leads attached to a finite-size sample, as exemplified by the bridge in Fig.~\ref{fig:setup}. 
This is analogous to profound developments in our understanding of quantum Hall effect ensuing 
from the comparison of the macroscopic charge transport in bulk samples with the gap in the 
energy spectrum to the mesoscopic transport through gapless chiral edge states of multiterminal bridges employed in experiments.~\cite{meso_hall}  

For example, within a finite-width strip no charge or spin current can flow across its boundaries, so that nonequilibrium spin accumulation~\cite{accumulation}  
will appear near the lateral edges to generate compensating current in the direction  
opposite to the spin Hall current. Very recent experiments~\cite{kato} have indeed demonstrated 
for the first time the existence of such spin accumulation, which has opposite sign on the 
lateral edges of the wire, as the manifestation of the spin Hall effect(s) in two-terminal 
devices. Thus, when ideal (i.e., spin and charge interaction free) transverse leads are attached 
at the lateral edges of the 2DEG region in Fig.~\ref{fig:setup}, pure ($I_2^\uparrow + I_2^\downarrow = 0$) spin current $I_2^s = \frac{\hbar}{2e} (I^{\uparrow}_2 - I^{\downarrow}_2)$  should emerge in the 
probe 2 of the bridge.

Here we predict a novel type of effect which exhibits the spin Hall 
phenomenology in {\em mesoscopic} finite-size structures. The transverse  pure spin 
current $I_2^s$  in Fig.~\ref{fig:setup} is induced  by injecting unpolarized charge current 
through the  longitudinal leads into the  SO coupled central region which contains no impurities. The central region is assumed here to  be a finite-size Rashba spin-split  2DEG, which is  {\em phase-coherent} (i.e., electron is described by a single wave function within the sample) and in the ballistic transport regime where  electrons do not feel any electric field while propagating through the 2DEG.  Within the 2DEG carriers  are subjected  to the Rashba type of SO coupling, which is described by the following effective mass Hamiltonian
\begin{equation}\label{eq:rashba}
\hat{H} = \frac{\hat{\bf p}^2}{2m^*}  + \frac{\alpha}{\hbar}\left(\hat{\bm \sigma}\times\hat{\bf p}\right) \cdot {\bf z} + V_{\rm conf}(y),
\end{equation}
that takes into account structure inversion asymmetry~\cite{rashba,rashba_review} (of 
the  confining electric potential and differing  band discontinuities at the heterostructure 
quantum  well  interface~\cite{winkler}). Here $\hat{\bf p}$ is the momentum operator in 2D 
space, $\hat{\bm \sigma} =(\hat{\sigma}_x,\hat{\sigma}_y,\hat{\sigma}_z)$ is the vector of the 
Pauli  spin matrices, $\alpha$ is  the strength of the Rashba SO coupling,~\cite{rashba,rashba_review,winkler} and $V_{\rm conf}(y)$ is the transverse confining potential.

The paper is organized as follows. In Sec.~\ref{sec:multiprobe} we recast the Landauer-B\" uttiker multiprobe charge current formulas in terms of spin-resolved charge current, which then allows us to introduce the multiprobe spin current formalism that yields the general expression for the linear  response 
spin Hall {\em conductance} of the bridge in Fig.~\ref{fig:setup}
\begin{equation} \label{eq:gh}
G_{sH} = \frac{\hbar}{2e} \frac{I_2^\uparrow - I_2^\downarrow}{V_1-V_4},
\end{equation}
in terms of the spin-resolved transmission probabilities between different leads. We apply this 
formalism to a perfectly clean bridge  to demonstrate  in Sec.~\ref{sec:ballistic} the existence of {\em three} non-zero spin conductances  (corresponding  to three components $[I_2^s]^x$, $[I_2^s]^y$, $[I_2^s]^z$ of the vector of pure spin current  flowing through  the transverse leads), two of 
which $G_{sH}^z = [I_2^s]^z/(V_1 -V_4)$ and  $G_{sH}^x = [I_2^s]^x/(V_1 -V_4)$ represent the 
signature of the mesoscopic spin Hall effect in Rashba spin-split structures. They are determined by  
the density of  electrons  and the Rashba SO coupling strength $\alpha$, as well as by the measuring geometry  of the whole device (i.e., interfaces, boundaries, and the attached electrodes). Furthermore, 
we find two different scaling laws for $G_{sH}^z$ and $G_{sH}^x$, depending on whether the device is smaller or greater than the {\em mesoscale} set by the spin precession length $L_{\rm SO}$ on which  spin precesses by an angle $\pi$. Section~\ref{sec:disorder} shows that the predicted effect is able to survive weak disorder (introduced as spin-independent scattering off static impurities)---the spin Hall conductances $G_{sH}^z$ and $G_{sH}^x$ gradually diminish from their maximum values (set in the clean limit of Sec.~\ref{sec:ballistic}) within the metallic diffusive regime and become negligible as the disorder is increased, but before the onset of strong localization effects is reached in phase-coherent SO coupled 2D structures.

The magnitude of the intrinsic  spin Hall effect in infinite homogeneous systems is captured by the spin Hall {\em conductivity} $\sigma_{sH} = j_y^z/E_x$ which relates pure spin current density $j_y^z$, flowing in the $y$-direction and carrying spins polarized solely along the $z$-axis, as a response to the longitudinal externally applied electric field $E_x$. Using our results from Sec.~\ref{sec:ballistic} and Sec.~\ref{sec:disorder}, we analyze in Sec.~\ref{sec:vs} if any quantitative connection can be established between the bulk spin Hall conductivity $\sigma_{sH}$ and the spin Hall conductance $G_{sH}^z$ (i.e., between the intrinsic spin Hall current density $j_y^z$, which is not conserved in the  bulk and depends on the whole SO coupled Fermi sea, and $[I_2^s]^z$ which we find to be  a Fermi-surface quantity and conserved total spin current throughout the ideal leads). This analysis reveals different origins of the mesoscopic spin Hall effect, which is governed by the processes on the mesoscale~\cite{so_force} $L_{\rm SO}$ and represents the nonequilibrium manifestation of SO couplings in confined ballistic semiconductor nanostructures. We conclude in Sec.~\ref{sec:conclusion}.

\section{Scattering approach to quantum transport of spin currents in multiprobe geometries}\label{sec:multiprobe}

The mesoscopic experiments on quantum Hall bridges in the early 1980s were posing 
a challenge for theoretical interpretation of multiterminal
transport measurements.~\cite{meso_hall} By viewing the current and voltage probes
on equal footing, B\" uttiker~\cite{buttiker} has provided an elegant solution to this problem in the form of a multiprobe formula~\cite{buttiker,baranger,datta_book} 
\begin{equation}\label{eq:buttiker}
I_p = \sum_q (G_{qp} V_p - G_{pq}V_q)  = \sum_q G_{pq}(V_p - V_q), 
\end{equation}
which relates, via the conductance coefficients $G_{pq}$, charge current $I_p=I_p^\uparrow + I_p^\downarrow$ in probe $p$ to the voltages $V_q$ in all other probes attached to the sample. To study 
the spin-resolved charge currents $I_p^\sigma$ ($\sigma=\uparrow,\downarrow$) of individual 
spin species $\uparrow$, $\downarrow$ we imagine that each non-magnetic lead in Fig.~\ref{fig:setup} consists of the two leads allowing only one spin species to propagate (as realized by, e.g., half-metallic ferromagnetic leads). Upon replacement $I_p \rightarrow I_p^\sigma$ and $G_{pq} \rightarrow G_{pq}^{\sigma \sigma^\prime}$, this viewpoint allows us to extract the multiprobe formulas for the spin-resolved charge currents~\cite{pareek,kiselev} $I_p^\sigma$,  thereby obtaining the linear response relation for spin current $I_p^s=\frac{\hbar}{2e} (I_p^\uparrow - I_p^\downarrow)$ flowing through the lead $p$
\begin{eqnarray} \label{eq:spinbuttiker}
I_p^s & = &
\frac{\hbar}{2e}\sum_q [ (G_{qp}^{\uparrow\uparrow} + G_{qp}^{\downarrow\uparrow} - G_{qp}^{\uparrow\downarrow} - G_{qp}^{\downarrow\downarrow}) V_p \nonumber \\
&& - (G_{pq}^{\uparrow\uparrow} + G_{pq}^{\uparrow\downarrow} - G_{pq}^{\downarrow\uparrow} - G_{pq}^{\downarrow\downarrow}) V_q ].
\end{eqnarray}
Below we simplify the notation by introducing the labels $G_{pq}^{\rm in}=G_{pq}^{\uparrow\uparrow}+G_{pq}^{\uparrow\downarrow}-G_{pq}^{\downarrow\uparrow}-G_{pq}^{\downarrow\downarrow}$ and $G_{pq}^{\rm out}=G_{pq}^{\uparrow\uparrow}+G_{pq}^{\downarrow\uparrow}-G_{pq}^{\uparrow\downarrow}-G_{pq}^{\downarrow\downarrow}$. Furthermore, these coefficients have transparent physical interpretation: $\frac{\hbar}{2e}G_{qp}^{\rm out} V_p$ is the spin current flowing from the lead $p$ with voltage $V_p$ into other leads $q$ whose voltages are $V_q$, while $\frac{\hbar}{2e} G_{pq}^{\rm in} V_q$ is the spin current flowing from the leads
$q \neq p$ into the lead $p$.

The standard charge conductance coefficients~\cite{buttiker,baranger,datta_book} in the multiprobe Landauer-B\" uttiker formalism  Eq.~(\ref{eq:buttiker}) are expressed in terms of the spin-resolved conductances as  $G_{pq}=G_{pq}^{\uparrow\uparrow}+G_{pq}^{\uparrow\downarrow}+G_{pq}^{\downarrow\uparrow}+G_{pq}^{\downarrow\downarrow}$. Their introduction in 1980s was prompted by the need to describe linear transport properties of a single sample, with specific impurity arrangements and attached to specific probe configuration, by using measurable quantities (instead of the bulk conductivity which is inapplicable to mesoscopic conductors~\cite{baranger}). They describe total charge current flowing  in and out of  the system 
in response to voltages applied  at its boundaries. 

Regardless of the detailed microscopic physics of transport, conductance coefficients must satisfy the sum rule $\sum_q G_{qp} = \sum_q G_{pq}$ in order to ensure the second equality in Eq.~(\ref{eq:buttiker}), i.e.,  the charge current must be zero $V_q={\rm const.} \Rightarrow I_p \equiv 0$ in equilibrium. On the other hand, the multiprobe spin current formulas Eq.~(\ref{eq:spinbuttiker}) apparently posses a nontrivial equilibrium solution $V_q={\rm const.} \Rightarrow I_p^s \neq 0$ (found in Ref.~\onlinecite{pareek}) that would be entirely alien to the Landauer-B\" uttiker paradigm demanding usage of only measurable quantities. However, when all leads are at the same potential, a purely equilibrium non-zero term $\frac{\hbar}{2e} (G_{pp}^{\rm out} V_p - G_{pp}^{\rm in} V_p) = \frac{\hbar}{e} (G^{\downarrow \uparrow}_{pp} - G^{\uparrow \downarrow}_{pp})V_p$  becomes relevant for $I_p^s$ [note that for devices in nonequilibrium the summation in Eq.~(\ref{eq:spinbuttiker}) goes only over $q \neq p$ leads], canceling all other terms in Eq.~(\ref{eq:spinbuttiker}) to ensure that no {\em unphysical} total spin current $I_p^s \neq 0$ can appear in the leads of an unbiased ($V_q$=const.) multiterminal device.~\cite{spin_hall_ring,kiselev_theorem}

At zero temperature, the spin-resolved conductance coefficients $G_{pq}^{\sigma \sigma^\prime}=\frac{e^2}{h} \sum_{ij} |{\bf t}^{pq}_{ij,\sigma \sigma^\prime}|^2$, where summation is over the conducting channels in the leads, are obtained from the Landauer-type formula as the probability for spin-$\sigma^\prime$  electron incident in lead $q$ to be transmitted to lead $p$  as spin-$\sigma$ electron. The quantum-mechanical probability amplitude for this processes is given by the matrix elements of the transmission matrix ${\bf t}^{pq}$, which is determined only by the wave functions (or Green functions) at the Fermi energy.~\cite{baranger} The stationary states of the structure 2DEG + two leads supporting one or two  conducting channels can be found exactly  by matching the wave functions in the leads to the eigenstates 
of the Hamiltonian Eq.~(\ref{eq:rashba}), thereby allowing one to obtain the  charge conductance from 
the Landauer transmission formula.~\cite{governale} However, modeling of the full bridge geometry with two extra leads  attached in the transverse  direction, as well as existence of many open conducting channels, requires  to switch from wave functions to some type of Green function formalism. 

For this purpose we represent the Rashba Hamiltonian~\cite{rashba_review,winkler} of the 2DEG in Fig.~\ref{fig:setup} in a local orbital basis defined on the $L \times L$ lattice (with lattice spacing $a$) as~\cite{purity}
\begin{eqnarray}\label{eq:tbh}
  \hat{H}  & = &  \left( \sum_{\bf m} \varepsilon_{\bf m}|{\bf m} \rangle \langle {\bf m}| 
  -   t_{\rm o} \sum_{\langle {\bf m},{\bf m}^\prime \rangle} |{\bf m} \rangle \langle {\bf m}^\prime|  \right) \otimes \hat{I}_s  \nonumber \\
  && + \frac{ \alpha}{\hbar} (\hat{p}_y \otimes \hat{\sigma}_x - \hat{p}_x \otimes
\hat{\sigma}_y).
\end{eqnarray}
Here $t_{\rm o}$ (a unit of energy) is the nearest-neighbor hopping between $s$-orbitals $\langle {\bf r}|{\bf m}\rangle = \psi({\bf r}-{\bf m})$ on adjacent atoms located at sites ${\bf m}=(m_x,m_y)$ of the lattice.  Since momentum operator in the tight-binding representation is $\langle {\bf m} |\hat{p}_x| {\bf m}^\prime \rangle = \delta_{m_x^\prime,m_x \pm 1} i \hbar \left( m_x - m^\prime_x \right)/2a^2$, the Rashba SO term  (in which $\otimes$ stands for the tensor product of operators)  introduces the SO hopping energy scale $t_{\rm SO}=\alpha/2a$. In a perfectly clean 2DEG the  on-site potential energy is $\varepsilon_{\bf m}=0$, while disordered 2DEG can be simulated via a random variable $\varepsilon_{\bf m} \in [-W/2,W/2]$ modeling short-range isotropic scattering off spin-independent impurities.

For non-interacting particle which propagates through a finite-size sample of arbitrary shape, the transmission matrices 
\begin{eqnarray} \label{eq:transmission}
{\bf t}^{pq} &  =  & \sqrt{-\text{Im} \, \hat{\Sigma}_p \otimes \hat{I}_s } \cdot \hat{G}^{r}_{pq} \cdot \sqrt{-\text{Im}\, \hat{\Sigma}_q \otimes \hat{I}_s}, \nonumber \\ 
\text{Im} \, \hat{\Sigma}_p &  = & \frac{1}{2i} \left( \hat{\Sigma}_p^r - \hat{\Sigma}_p^a \right), 
\end{eqnarray}
between different leads can be evaluated in a numerically exact fashion using  the real$\otimes$spin-space Green functions.~\cite{purity} This requires to compute a single object, the retarded Green operator 
\begin{equation}\label{eq:green}
\hat{G}^r = \frac{1}{E\hat{I}_o \otimes \hat{I}_s-\hat{H} - \sum_{p=1}^4 \hat{\Sigma}^r_p \otimes \hat{I}_s},
\end{equation} 
which becomes a matrix (i.e., the Green function)  when represented in a basis $|{\bf m} \rangle \otimes |\sigma \rangle \in {\mathcal H}_o \otimes {\mathcal H}_s$ introduced by the Hamiltonian Eq.~(\ref{eq:tbh}). Here $|\sigma \rangle$ are the eigenstates of the spin operator for the chosen spin quantization axis. 
The matrix elements $G^r({\bf m}^\prime \sigma^\prime;{\bf m},\sigma) = \langle {\bf m}^\prime,\sigma^\prime|\hat{G}^r| {\bf m},\sigma\rangle$  yield the probability amplitude for an electron to propagate between two arbitrary locations ${\bf m}$ and ${\bf m}^\prime$ (with or without flipping its spin $\sigma$ during the motion) inside an open conductor in the absence of inelastic processes. Its submatrix $\hat{G}^{r}_{pq}$, which is required in Eq.~(\ref{eq:transmission}), consists of those matrix elements which connects the layer of the sample attached to the lead $p$ to the layer of the sample attached to the lead $q$. 
The unit operators $\hat{I}_o$ 
and $\hat{I}_s$ act in the orbital ${\mathcal H}_o$ and the spin Hilbert spaces ${\mathcal H}_s$, respectively, which comprise the Hilbert space of a single spinfull particle  ${\mathcal H}_o \otimes {\mathcal H}_s$ (via tensor product of vector spaces). The self-energy $\sum_{p=1}^4 \hat{\Sigma}^r_p \otimes \hat{I}_s$  ($r$-retarded, $a$-advanced, $\hat{\Sigma}_p^{a}=[\hat{\Sigma}_p^{r}]^{\dagger}$) 
account  for the ``interaction'' of an open system with the attached four ideal semi-infinite leads~\cite{datta_book} $p$.

A direct correspondence between  the continuous effective Rashba Hamiltonian Eq.~(\ref{eq:rashba}) [with parabolic  energy-momentum dispersion] and its lattice version Eq.~(\ref{eq:tbh}) [with tight-binding dispersion]  is established  by selecting $E_F$ close to the bottom of the band (where tight-binding dispersion reduces to the quadratic one), and by using $t_{\rm o}=\hbar^2/(2 m^* a^2)$ for the orbital hopping  which yields the effective mass $m^*$ in the continuum limit. We elucidate further the connection between the standard effective Rashba Hamiltonian in continuous representation Eq.~(\ref{eq:rashba}) and its lattice version  by interpreting the tight-binding parameters in Eq.~(\ref{eq:tbh}) for particular experimental realization of a 2DEG in semiconductor heterostructures.  For example, the InGaAs/InAlAs heterostructure employed in experiments of Ref.~\onlinecite{nitta}  is characterized by  the effective mass $m^*=0.05m_0$ ($m_0$ is  the free electron mass) and the  width of the conduction band $\Delta=0.9$ eV, which sets $t_{\rm o}=\Delta/8=0.112$ meV for the  orbital hopping parameter on a square lattice (with four nearest neighbors of each site) and $a \simeq 2.6$ nm for its  lattice spacing. Thus, the Rashba SO coupling of 2DEG formed in this  heterostructure, tuned to a maximum value~\cite{nitta} $\alpha=0.93 \cdot 10^{-11}$ eVm by the gate voltage covering the 2DEG, corresponds to the SO hopping $t_{\rm SO}/t_{\rm o} \simeq 0.016$ in the lattice Hamiltonian Eq.~(\ref{eq:tbh}).

\subsection{General expression for the spin Hall conductance}\label{sec:general}

Since the total charge current $I_p=I_p^\uparrow +
I_p^\downarrow$ depends only on the voltage difference between the
leads in Fig.~\ref{fig:setup}, we set one of them to zero (e.g.,
$V_4=0$ is chosen as the reference potential) and apply voltage $V_1$ to the structure. Imposing the requirement $I_2 = I_3 =0$  for the  voltage probes 2 and 3 allows us to get the voltages $V_2/V_1$ and $V_3/V_1$ 
by inverting the multiprobe charge current formulas Eq.~(\ref{eq:buttiker}). Finally, by 
solving Eq.~(\ref{eq:spinbuttiker}) for $I_2^s$ we obtain the most general expression for 
the spin Hall conductance defined by Eq.~(\ref{eq:gh})
\begin{equation} \label{eq:gh_explicit}
G_{sH}=\frac{\hbar}{2e} \left[(G_{12}^{\rm out}+G_{32}^{\rm out}+ G_{42}^{\rm out})\frac{V_2}{V_1} - G_{23}^{\rm in}\frac{V_3}{V_1} - G_{21}^{\rm in} \right].
\end{equation}
This quantity is measured in the units of the spin conductance quantum $e/4\pi$ (as the largest possible  $G_{sH}$ in the transverse leads with only one open conducting channel), which is  the counterpart of a familiar charge conductance quantum $e^2/h$ (as the natural unit for spin-resolved conductance coefficients $G_{pq}^{\sigma \sigma^\prime}$).

In contrast to the charge current which is a scalar quantity, spin current has three components because of the vector nature of spin (i.e., different ``directions'' of spin correspond to different  quantum mechanical superpositions of $|\!\! \uparrow \rangle$ and $|\!\! \downarrow \rangle$  states). Therefore, we can expect that, in general, the detection of spin transported through the transverse leads of mesoscopic devices will find its expectation values to be non-zero for all three axes. We indeed find in Sec.~\ref{sec:ballistic} and Sec.~\ref{sec:disorder} that all three components of the spin current in the transverse leads 2 and 3 are non-zero. However, their flow 
properties
\begin{subequations}\label{eq:flow}
\begin{eqnarray}
\lbrack I_2^s \rbrack^z & = & -\lbrack I_3^s \rbrack^z,\\  
\lbrack I_2^s \rbrack^x & =& -\lbrack I_3^s \rbrack^x, \\  
\lbrack I_2^s \rbrack^y & = & \lbrack I_3^s \rbrack^y,
\end{eqnarray}
\end{subequations}
show that only the $z$- and the $x$-components represent the spin Hall response for the Rashba SO coupled  four-terminal bridges. That is, if we connect the transverse leads 2 and 3 to each other (thereby connecting the lateral edges of 2DEG by a wire), only the spin current carrying $z$- and $x$-polarized spins will flow through them, as expected from the general spin  Hall phenomenology where nonequilibrium spin Hall accumulation detected in experiments~\cite{kato} has opposite sign~\cite{accumulation} on the lateral edges 
of 2DEG.

Therefore, to quantify all non-zero components of the vector of transverse spin current in the linear response regime, we introduce three spin conductances $G_{sH}^x=[I_2^s]^x/V_1$, $G_{sp}^y=[I_2^s]^y/V_1$, and $G_{sH}^z=[I_2^s]^z/V_1$ (assuming $V_4=0$). They can be evaluated using the same general formula Eq.~(\ref{eq:gh_explicit}) where the spin quantization axis for $\uparrow$, $\downarrow$ in spin-resolved charge conductance coefficients is chosen to be the $x$-, $y$-, or $z$-axis, respectively. For example,  selecting  $\hat{\sigma}_z  |\!\! \uparrow \rangle = + |\!\! \uparrow \rangle$ and $\hat{\sigma}_z |\!\! \downarrow \rangle=  - |\!\! \downarrow \rangle$ for the basis in which the Green operator Eq.~(\ref{eq:green}) is represented allows one to compute the $z$-component of  the spin current $[I_p^s]^z$. In accord with 
their origin revealed by Eq.~(\ref{eq:flow}), we denote $G_{sH}^z$ and $G_{sH}^x$ as the spin Hall conductances, while $G_{sp}^y$ is labeled as the ``spin polarization'' conductance since 
it stems from the polarization of 2DEG by the flow of unpolarized charge current in the presence of SO couplings~\cite{accumulation,ganichev} (see also Sec.~\ref{sec:ballistic}).

\subsection{Symmetry properties of  spin conductances}\label{sec:symmetry}

Symmetry properties of the conductance coefficients with respect to the reversal of a bias 
voltage or the direction of an external magnetic field play an essential role in our 
understanding of linear response  electron transport in macroscopic and mesoscopic conductors.~\cite{buttiker,datta_book} For example, in the absence of magnetic field they satisfy  $G_{pq}=G_{qp}$ (which can be proved assuming a particular model for charge transport~\cite{datta_book}). Moreover, since the effective magnetic field ${\bf B}_R({\bf p})$  of the Rashba SO coupling depends on momentum, it does not break the time-reversal invariance which imposes the following 
property on the spin-resolved conductance coefficients $G_{pq}^{\sigma \sigma^\prime} = G_{qp}^{- \sigma^\prime -\sigma}$ in the multiterminal SO coupled bridges.~\cite{pareek,kiselev}

In addition, the ballistic four-terminal bridge in Fig.~\ref{fig:setup} with no impurities posseses various geometrical symmetries. It is invariant under rotations and reflections that interchange the leads, such as:  (i) rotation $C_4$ ($C_2$) by an angle $\pi/2$ ($\pi$) around the $z$-axis for a square (rectangular) 2DEG central region; (ii) reflection $\sigma_{vx}$ in the $xz$-plane; and (iii) reflection $\sigma_{vy}$ in the $yz$-plane. These geometrical symmetries, together with $G_{pq}=G_{qp}$ property, specify $V_2/V_1 = V_3/V_1 \equiv 0.5$ solution for the voltages of the transverse leads when $I_2 = I_3=0$  condition is imposed on their currents.  

The device Hamiltonian containing the Rashba SO term commutes with the unitary transformations which represent these symmetry operations in the Hilbert space  ${\mathcal H}_o \otimes {\mathcal H}_s$: (i) $\hat{U}(C_2) \otimes \exp{(i\frac{\pi}{2}\hat{\sigma}_z)}$, which performs the transformation $\hat{\sigma}_x \rightarrow -\hat{\sigma}_x$,  $\hat{\sigma}_y \rightarrow - \hat{\sigma}_y$, $\hat{\sigma}_z \rightarrow \hat{\sigma}_z$ and interchanges the leads 1 and 4 as well as the leads 2 and 3; (ii) $\hat{U}(\sigma_{vx}) \otimes\exp{(i\frac{\pi}{2}\hat{\sigma}_y)}$, which transforms the Pauli matrices  $\hat{\sigma}_x \rightarrow -\hat{\sigma}_x$,  $\hat{\sigma}_y \rightarrow \hat{\sigma}_y$, $\hat{\sigma}_z \rightarrow - \hat{\sigma}_z$ and interchanges leads 2 and 3; and  (iii) $\hat{U}(\sigma_{vy}) \otimes\exp{(i\frac{\pi}{2}\hat{\sigma}_y)}$ which transforms  $\hat{\sigma}_x \rightarrow \hat{\sigma}_x$,  $\hat{\sigma}_y \rightarrow -\hat{\sigma}_y$,  $\hat{\sigma}_z \rightarrow -\hat{\sigma}_z$ and exchanges lead 1 with lead  4. The Hamiltonian also commutes with the time-reversal operator $K \exp{(i\frac{\pi}{2}\hat{\sigma}_y)}$ [where $K$ is the complex conjugation operator].

The effect of  these symmetries on the spin-resolved charge conductance coefficients, and  the 
corresponding spin conductances $G_{sH}^z$, $G_{sH}^x$ and $G_{sp}^y$ expressed in terms of them 
through Eq.~(\ref{eq:gh_explicit}), is as follows. The change in the sign of the spin operator means that  spin-$\uparrow$ becomes  spin-$\downarrow$ so that, e.g.,  $G_{pq}^{\rm in}$ will be transformed into $-G_{qp}^{\rm in}$. Also, the time reversal implies changing the signs of all spin operators and all momenta so that $G_{pq}^{\sigma \sigma^\prime} = G_{qp}^{- \sigma^\prime -\sigma}$  is equivalent to $G_{pq}^{\rm in}=-G_{qp}^{\rm out}$. Thus, invariance with  respect to $\hat{U}(\sigma_{vy}) \otimes\exp{(i\frac{\pi}{2}\hat{\sigma}_y)}$ yields the identities  $G_{21}^{{\rm in},x}  =  G_{24}^{{\rm in},x}$, $G_{21}^{{\rm in},y} = -G_{24}^{{\rm in},y}$, and $G_{21}^{{\rm in},z}  =  -G_{24}^{{\rm in},z}$. These symmetries do not imply cancellation of $G_{23}^{{\rm in},x}$. However, invariance with respect to $\hat{U}(\sigma_{vx}) \otimes\exp{(i\frac{\pi}{2}\hat{\sigma}_y)}$ and $\hat{U}(C_4) \otimes \exp{(i\frac{\pi}{4}\hat{\sigma}_z)}$ implies that $G_{23}^{{\rm in},y} \equiv 0$ and $G_{23}^{{\rm in},z} \equiv 0$. 

These symmetry imposed conditions simplify the general formula Eq.~(\ref{eq:gh_explicit}) for spin conductances of a perfectly clean Rashba SO coupled four-terminal bridge to
\begin{subequations} \label{eq:simple_rashba}
\begin{eqnarray}
G_{sH}^{x} & = & 2G_{12}^{{\rm out},x}+G_{32}^{{\rm out},x}, \\
G_{sH}^{y} & = & G_{12}^{{\rm out},y}, \\
G_{sH}^{z} & = & G_{12}^{{\rm out},z}.
\end{eqnarray}
\end{subequations}
where we employ the result $V_2/V_1 = V_3/V_1 \equiv 0.5$ valid for a geometrically symmetric clean bridge. Because this solution for the transverse terminal voltages is violated in disordered bridges, 
its sample specific (for given impurity configuration) spin conductances cannot be computed from simplified formulas Eq.~(\ref{eq:simple_rashba}).

It insightful to apply the same symmetry analysis to the bridges with other types of SO couplings. 
For example, if the Rashba term in the Hamiltonian Eq.~(\ref{eq:rashba}) is replaced by the linear Dresselhaus SO term $\frac{\beta}{\hbar} \left(\hat{p}_x \hat{\sigma}_x  - \hat{p}_y \hat{\sigma}_y  \right)$ due to 
bulk inversion asymmetry,~\cite{spintronics} no qualitative change in our analysis ensues since the two SO couplings can be transformed into each  other by a unitary matrix $(\hat{\sigma}_x + \hat{\sigma}_y)/\sqrt{2}$. In this case, the spin Hall response is signified by  $[I_2^s]^z  =  -[I_3^s]^z$ and $[I_2^s]^y = -[I_3^s]^y$ components of the transverse spin current, while $[I_2^s]^x = [I_3^s]^x$. For the Dresselhaus SO coupled bridge, the general expression Eq.~(\ref{eq:gh_explicit}) simplifies to
\begin{subequations}
\begin{eqnarray}\label{eq:simple_dresel}
G_{sH}^{x} & = & G_{12}^{{\rm out},x}, \\
G_{sH}^{y} & = & 2G_{12}^{{\rm out},y}+G_{32}^{{\rm out},y}, \\
G_{sH}^{z} & = & G_{12}^{{\rm out},z}. 
\end{eqnarray}
\end{subequations}
The qualitatively different situation emerges when both the Rashba and the linear Dresselhaus SO couplings become  relevant in the central region of the bridge since in this case it is impossible to find spin rotation which, combined with the spatial symmetry, would keep the Hamiltonian invariant while only transforming the signs of its spin matrices. Moreover, for such ballistic bridge the condition  condition $I_2 = I_3 = 0$  leads to $V_2/V_1 = 1 - V_3/V_1$ solution for the voltages, whereas imposing the alternative condition $V_2=V_3$  generates non-zero charge currents flowing through the transverse leads 2 and 3 together with the spin currents [for which no simple relations akin to Eq.~(\ref{eq:flow}) can be written in either of these cases].

\section{Transverse pure spin currents in ballistic bridges}\label{sec:ballistic}
 
\begin{figure}
\centerline{\psfig{file=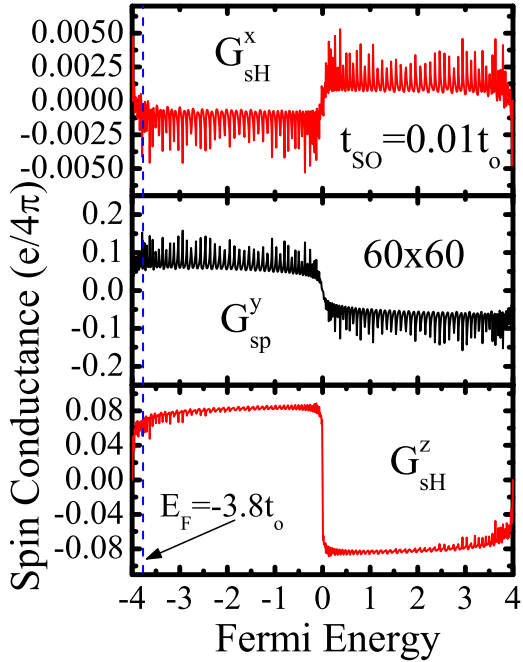,width=2.85in,angle=0} }
\caption{(Color online) The Fermi energy dependence of the spin Hall $G_{sH}^z$, $G_{sH}^x$ and spin polarization  $G_{sp}^y$ conductances in ballistic four-probe bridges (Fig.~\ref{fig:setup}) where the central region is 2DEG of the size $60a \times 60a$. Within the 2DEG electrons are subjected to the Rashba SO coupling whose strength is $t_{\rm SO}=\alpha/2a=0.01t_{\rm o}$.} \label{fig:gh_fermi}
\end{figure}

\begin{figure}
\centerline{\psfig{file=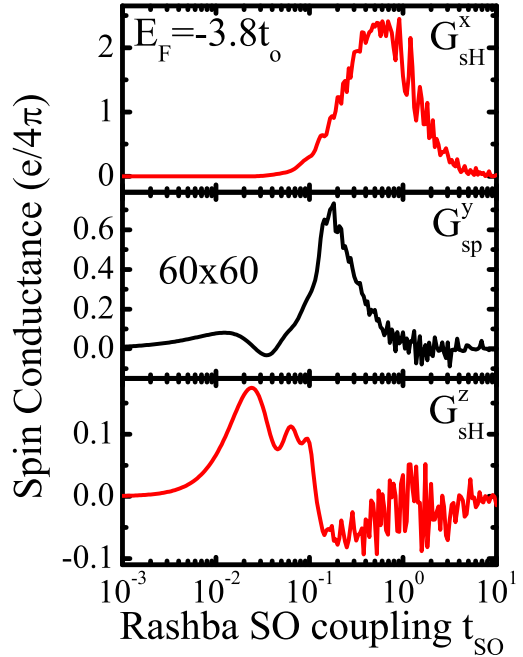,width=2.8in,angle=0} }
\caption{(Color online) The dependence of the spin Hall $G_{sH}^z$, $G_{sH}^x$ and spin polarization  $G_{sp}^y$  conductances on the Rashba SO coupling strength $t_{\rm SO}$ in the 2DEG of the size $60a \times 60a$ attached to four ideal (with no spin $t_{\rm SO} = 0$ and charge interactions) leads. The
unpolarized current injected through the longitudinal leads is composed of spin-$\uparrow$ and spin-$\downarrow$ electrons at the Fermi energy $E_F=-3.8t_{\rm o}$.} \label{fig:gh_rashba}
\end{figure}

\begin{figure*}
\centerline{\psfig{file=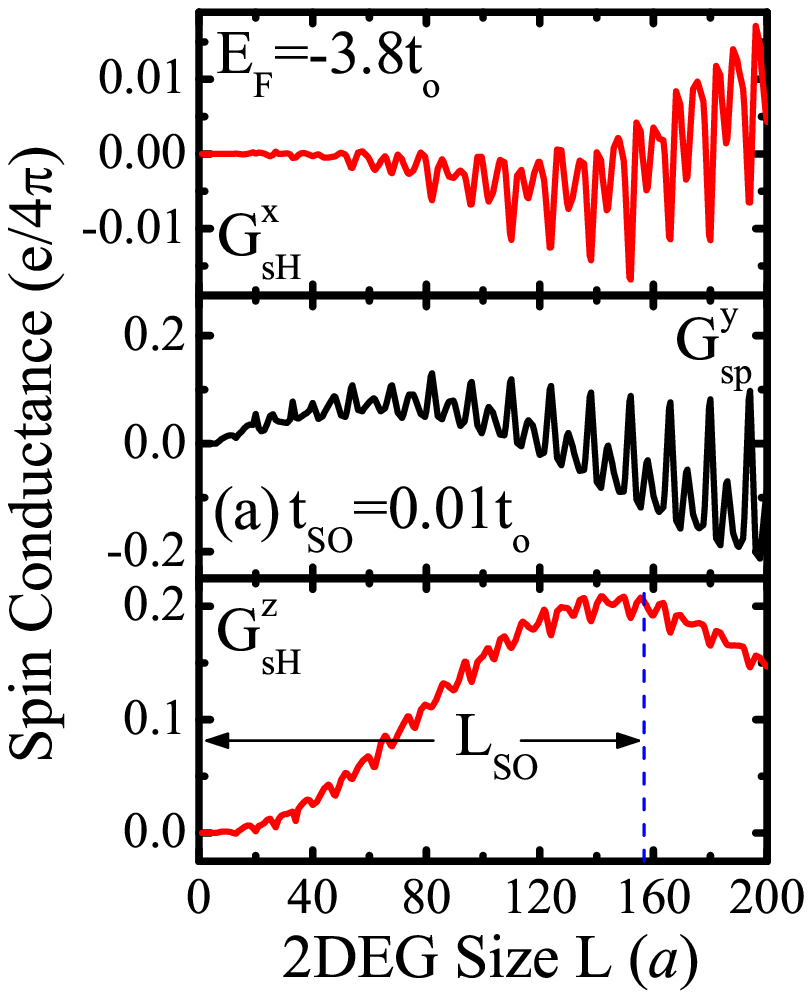,width=2.8in,angle=0} \hspace{0.5in} \psfig{file=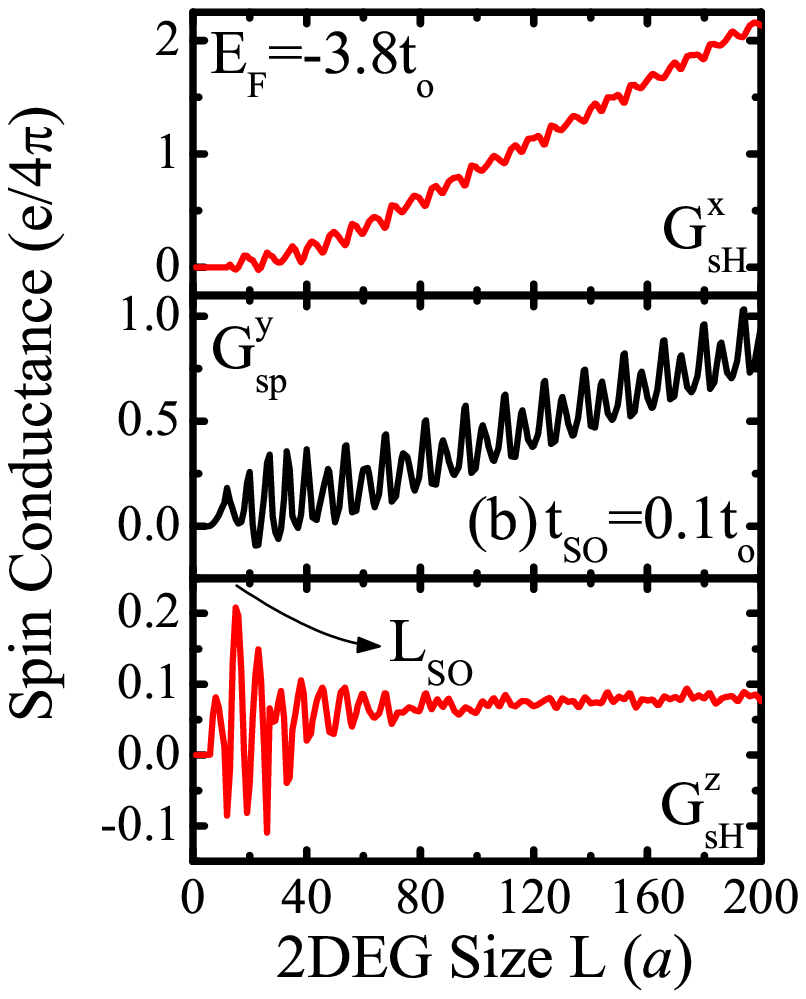,width=2.8in,angle=0} }
\caption{(Color online) The finite-size scaling of the spin Hall $G_{sH}^z$, $G_{sH}^x$ and spin polarization $G_{sp}^y$ conductances in ballistic four-probe bridges with the Rashba SO coupling strength: (a) $t_{\rm SO}=0.01t_{\rm o}$; and (b) $t_{\rm SO}=0.1t_{\rm o}$. The corresponding spin precession length denoted on the graphs is: (a) $L_{\rm SO} \approx 157a$; and (b) $L_{\rm SO} \approx 15.7a$.}\label{fig:gh_size}
\end{figure*}

Figures~\ref{fig:gh_fermi}, ~\ref{fig:gh_rashba}, and ~\ref{fig:gh_size} demonstrate that the 
spin Hall conductance is not universal---$G_{sH}(E_F,t_{\rm SO},L,W)$ depends on the 2DEG 
parameters  such as the  density of charge carriers (i.e., the Fermi energy $E_F$), the Rashba SO 
coupling $t_{\rm SO}=\alpha/2a$, and the system size $L$. Furthermore, due 
to the sensitivity of spin dynamics in confined SO coupled ballistic systems to the boundaries 
and interfaces,~\cite{chao} it can also be affected by the measuring geometry. 


Nevertheless, we find in Fig.~\ref{fig:gh_rashba} that all square 2DEG samples of the size $L_{\rm SO} \times L_{\rm SO}$, where $L_{\rm SO}$ is the spin precession length, exhibit approximately the same  $G_{sH}^z \simeq 0.2 e/4\pi$. Therefore, we pay special attention to the intertwined effect of $\alpha$ and $L$ brought about by the fact that SO couplings introduce a characteristic length scale into the bridge---the spin precession  length $L_{\rm SO} = \pi/ 2 k_{\rm SO}$ over which spin precesses by an angle $\pi$ (i.e., the state  $|\!\! \uparrow \rangle$ evolves into $|\!\! \downarrow \rangle$). In the case of the Rashba SO coupling, $2k_{\rm SO} = 2m^*\alpha/\hbar^2$ is  the  difference of Fermi wave vectors for the spin-split  transverse energy subbands of a quantum  wire. This quantity is the same for all subbands of the quantum wire in the case of  parabolic  energy-momentum dispersion so that single parameter 
\begin{equation}\label{eq:solength}
L_{\rm SO} = \frac{\pi t_{\rm o}}{2t_{\rm SO}} a
\end{equation}  
characterizes the whole structure. For example, the mesoscale $L_{\rm SO}$ sets the characteristic 
length for the evolution of the nonequilibrium spin polarization in the course of transport through 
the ballistic,~\cite{chao} as well as  the diffusive SO coupled structures (which are sufficiently wide 
and  weakly disordered~\cite{purity}) where it plays  the role of the disorder-independent  
D'yakonov-Perel' spin relaxation length.~\cite{spintronics,chao,wees} 

In this section we study systematically the  dependence of the spin Hall and spin polarization conductances  on the first three basic parameters of the bridge, while leaving the effect of the disorder strength  $W$ for Sec.~\ref{sec:disorder} (cf. Fig.~\ref{fig:gh_disorder}), and the influence of the measuring geometry determined by  the attached leads to Sec.~\ref{sec:vs} (cf. Fig.~\ref{fig:gh_adiabatic}).

\subsection{Fermi energy dependence of the spin Hall conductance}

The spin conductances plotted in Fig.~\ref{fig:gh_fermi} are an odd function of the Fermi energy and, therefore,  have to vanish $G_{sH}^{z,x}(E_F=0) \equiv 0$, $G_{sp}^y(E_F=0) \equiv 0$ at the half-filled band $E_F=0$. This feature is a consequence of the spin current being defined as the difference  of charge currents of spin-$\uparrow$ and spin-$\downarrow$ electrons and particle-hole symmetry of the tight-binding Hamiltonian Eq.~(\ref{eq:tbh}). That is, the  spin current carried by electrons above the half-filling $E_F>0$ can be interpreted as the propagation of positively charged holes which move in the opposite direction, and have opposite  spin, to that of electrons. To evade artifacts of the tight-binding energy-momentum  dispersion, which can enhance the spin Hall conductance as we approach the band center, we highlight in Fig.~\ref{fig:gh_fermi} the values of spin conductances at the Fermi energy $E_F=-3.8t_{\rm o}$ chosen 
for our subsequent analysis. When zero-temperature unpolarized charge quantum transport is determined by 
the states at this $E_F$, which is close to the bottom of the band $E_b=-4t_{\rm o}$, the injected quasiparticles have quadratic and isotropic energy-momentum dispersion which characterizes the  Hamiltonians in effective mass approximation, such as the Rashba one in Eq.~(\ref{eq:rashba}). Also the Fermi wavelength corresponding to the Fermi energy measured from the band bottom $E_F - E_b = 0.2t_{\rm o}$ is much greater than the lattice spacing $a$ so that possible artifacts of the discretization are avoided.

\subsection{Rashba SO coupling dependence of the spin Hall conductances}

Considerable interest for exploiting the Rashba SO coupling~\cite{rashba,rashba_review,winkler} for  semiconductor spintronics applications stems from the possibility to tune its strength via an external gate electrode,~\cite{nitta} thereby manipulating spin solely by electrical means. A surprising early result in the theory of the intrinsic spin Hall effect is  apparent `universality' of $\sigma_{sH} = e/8\pi$ (obtained from the linear response theory for the clean Rashba Hamiltonian of an infinite 2DEG) in the sense that it does not depend on the strength of the SO coupling.~\cite{sinova} However, when scattering of impurities is taken into account in the limit $\alpha \rightarrow 0$, one recovers~\cite{loss} the physically expected result $\lim_{\alpha \rightarrow 0} \sigma_{sH} \rightarrow 0$.

Our exact treatment of transport of non-interacting quasiparticles through a clean finite-size system does 
not face any technical impediments in locating the lower limit on the strength of SO coupling capable of inducing the non-negligible spin Hall conductance $G_{sH}^z$, as shown in Fig.~\ref{fig:gh_rashba}. Although 
realistic Rashba SO coupling strengths in current experiments can be tuned within  the range 
$0.01t_{\rm o} \lesssim t_{\rm SO} \lesssim 0.1t_{\rm o}$ (see Sec.~\ref{sec:multiprobe}), we also 
show in the same Figure the upper limit of large values of $t_{\rm SO}$ beyond which all three 
components of the transverse spin current $I_2^s$ vanish due to the carrier reflection~\cite{purity,governale} at the interface  between the ideal lead with $t_{\rm SO} \equiv 0$ and the sample with strong SO coupling $t_{\rm SO} \sim t_{\rm o}$. The maximum $G_{sH}^z(t_{\rm SO})$ for square shaped 2DEGs attached to ideal semi-infinite leads is  obtained for samples of the size $L_{\rm SO} \times L_{\rm SO}$, which connects  
this dependence $G_{sH}^z(t_{\rm SO})$ with the finite-size  scaling properties of $G_{sH}^z(L)$ discussed 
in the next section.

\subsection{Finite-size scaling of the spin Hall conductances} \label{sec:scaling}

Figure~\ref{fig:gh_rashba} emphasizes the importance of the spin precession length scale for the spin Hall effect in ballistic semiconductor nanostructures. That is, the spin Hall conductance $G_{sH}^z(L)$ is increasing non-monotonically with the system size for $L < L_{\rm SO}$, attaining the maximum value $G_{sH}^z(L_{\rm SO}) \simeq 0.2 e/4\pi$. In such $L \times L$ samples with $L < L_{\rm SO}$, the other spin Hall conductance $G_{sH}^x(L)$ is negligible. The emergence of $[I_2^s]^x$ component of the spin Hall current can be understood heuristically by invoking the semiclassical picture involving the transverse SO force operator~\cite{so_force}
\begin{equation}\label{eq:so_force}
\hat{\bf F}_{\rm SO} = \frac{2 \alpha^2 m^* }{\hbar^3}  (\hat{\bf p} \times {\bf
z}) \otimes \hat{\sigma}^z  - \frac{d V_{\rm conf}(\hat{y})}{d\hat{y}} {\bf y}.
\end{equation}
and studying its effect on the propagation of spin-polarized wave packets.~\cite{so_force} This spin-dependent terms here are generated by the Rashba SO coupling term in the single-particle Hamiltonian of a clean 2DEG.  Its  expectation values in the  wave packets $|\Psi \rangle \otimes |\sigma \rangle$ (spin-polarized along the $z$-axis,  $\hat{\sigma}_z  |\!\! \uparrow \rangle = + |\!\! \uparrow \rangle$ and $\hat{\sigma}_z |\!\! \downarrow \rangle=  - |\!\! \downarrow \rangle$) shows that spin-$\uparrow$ and spin-$\downarrow$ injected electrons will be deflected in opposite transverse directions (e.g., spin-$\uparrow$ is initially~\cite{so_force} deflected to the right).  Moreover, their spins are forced into precession since injected $|\!\! \uparrow \rangle$, $|\!\! \downarrow \rangle$ states are not the eigenstates of the Zeeman term  $\hat{\bm \sigma} \cdot {\bf B}_R({\bf p})$, where the Rashba effective magnetic field ${\bf B}_R({\bf p})$ remains nearly  parallel to the $y$-axis due to the transverse confining potential~\cite{governale,purity} $V_{\rm conf}(y)$. 

The precession of the deflected spins is responsible for the oscillatory character~\cite{so_force} of the SO ``force'' (viewed as the expectation value of the SO force operator in the spin-polarized wave packet states) which changes sign along the wire. Thus, such $\alpha^2$-dependent spin-deflecting ``force'' can induce the change in the sign of the spin Hall current as a function of the system size, as shown in Fig.~\ref{fig:gh_rashba}(b) for strong SO coupling $t_{\rm SO}=0.1t_{\rm o}$. In the case of weaker SO couplings $t_{\rm SO} \lesssim  0.04t_{\rm o}$, the spin Hall conductance $G_{sH}^z$ oscillates for $L \lesssim L_{\rm SO}$ while remaining  positive, as exemplified by Fig.~\ref{fig:gh_rashba}(a). Note that in the convention of multiprobe spin current formulas Eq.~(\ref{eq:spinbuttiker}), positive spin Hall conductance means that spin current $I_2^s$ is flowing out of the lead 2 because spin-$\uparrow$ electrons are deflected to the right (i.e., toward the electrode 3 if injected from the electrode 1) and spin-$\downarrow$ electrons are deflected to the left, as expected from Eq.~(\ref{eq:so_force}).

The semiclassical picture of the SO ``force'' also explains the symmetry properties of the spin Hall current $[I_2^s]^z(-V)=[I_2^s]^z(V)$ with respect to the voltage bias reversal [i.e., the reversal of the momentum in Eq.~(\ref{eq:so_force})] or the reversal of the sign of the Rashba SO coupling, $[I_2^s]^z(-\alpha)=[I_2^s]^z(\alpha)$. These two features make it possible to differentiate~\cite{accumulation,so_force} between 
the $z$- and the $x$-component of the spin Hall current in the transverse leads since $[I_2^s]^x(-V)=[I_2^s]^x(V)$ and  $[I_2^s]^x(-\alpha)=-[I_2^s]^x(\alpha)$, which stem from the 
properties of the effective magnetic field ${\bf B}_R({\bf p})$ inducing the spin precession 
under these transformations.

The $y$-component of the transverse spin current satisfies $[I_2^s]^y=[I_3^s]^y$,  signaling  
that $G_{sp}^y$ is of completely different origin. It stems from  $\langle S_y({\bf r}) \rangle \neq 0$ nonequilibrium spin accumulation, which has the same sign on the lateral edges of 2DEG,~\cite{accumulation} induced when unpolarized charge current is flowing through the Rashba SO coupled 2DEG attached to two electrodes.~\cite{accumulation,ganichev,shytov} Thus, when transverse leads are connected to  the lateral edges the 2DEG, the spin-dependent chemical potential~\cite{wees} on the edges will push the spin  current $[I_2^s]^y$ into the leads. Since this chemical potential is the same on both lateral edges, connecting the edges by a wire would not lead to any net spin flux through it, in sharp contrast to the currents 
$[I_2^s]^z$ and $[I_3^s]^x$ that will transport spin through such wire as a signature of the spin Hall effect(s) phenomenology.

\section{Transverse pure spin currents in disordered bridges}\label{sec:disorder}

The most conspicuous difference between  the ``old'' extrinsic~\cite{extrinsic_1,extrinsic_2} 
effect (in paramagnetic metals or semiconductors without SO splitting of quasiparticle energies) and 
the ``new''  intrinsic spin Hall effect (in semiconductors with sufficiently large  SO splitting of quasiparticle energies) is that the former  vanishes in the ballistic limit, while the later persists even when no skew-scattering at impurities takes place.~\cite{murakami,sinova} However, the distinction between 
the two spin Hall effects turns out to be  ambiguous~\cite{loss,sugimoto} when SO energy splitting $\alpha k_F$ ($k_F$ is the Fermi  wave vector) is smaller than the disorder induced broadening of the energy levels $\hbar/\tau$ (where $\tau$ is the transport lifetime).  

Moreover, the resilience of the  intrinsic effect to scattering off static impurities has become a major issue in current debates over  the observability of spin Hall current in realistic samples of 2DEG with the Rashba SO interaction~\cite{shytov,loss,sugimoto,inoue,chalaev} or in hole-doped bulk 3D semiconductors.~\cite{murakami2004,loss} Early (lowest order) perturbative treatment of the semiclassical Boltzmann diffusive transport  in infinite homogeneous Rashba spin-split  2DEG has revealed that $\sigma_{sH}$ could survive disorder effects with the proviso that $\alpha k_F \tau/\hbar \gg 1$, while  being gradually reduced from the `universal' value $e/8\pi$ with increasing disorder strength.~\cite{loss}  However, when vertex corrections are included in the perturbative expansion, the intrinsic effect  turns out to be suppressed $\sigma_{sH} \rightarrow 0$  at {\em arbitrarily} small disorder $\alpha k_F \tau/\hbar \rightarrow \infty$, as revealed by a multitude of transport approaches~\cite{inoue} applied to any system with linear in momentum SO energy splitting (note that such cancellation induced by the ladder vertex corrections does not occur when SO coupling contains higher-order momentum terms~\cite{murakami2004}). 

The vanishing intrinsic spin Hall current density $j_y^z \rightarrow 0$ is also found at the weak localization level in the perturbative expansion in small parameter~\cite{chalaev} $1/k_F\ell$, as well as in the bulk of 2DEG (infinite  in  the transverse direction) attached to two massive electrodes in the longitudinal direction where, nevertheless, macroscopic inhomogeneities can induce non-zero $j_y^z \neq 0$ within the spin relaxation length $L_{\rm SO}$ wide region around the electrode-2DEG interfaces.~\cite{shytov}  Finally, recent reexamination~\cite{sugimoto} of these results in a vast range of ratios of the SO coupling induced $\alpha k_F$ and the disorder induced $\hbar/\tau$ energy scales confirms that $\sigma_{sH} \rightarrow 0$ is indeed suppressed in both the clean $\alpha k_F \tau/\hbar \rightarrow \infty$ and dirty limits $\alpha k_F \tau/\hbar \rightarrow 0$, while being able to attain a non-zero optimum value at $\alpha k_F \tau/\hbar \simeq 10$.

\begin{figure}
\centerline{\psfig{file=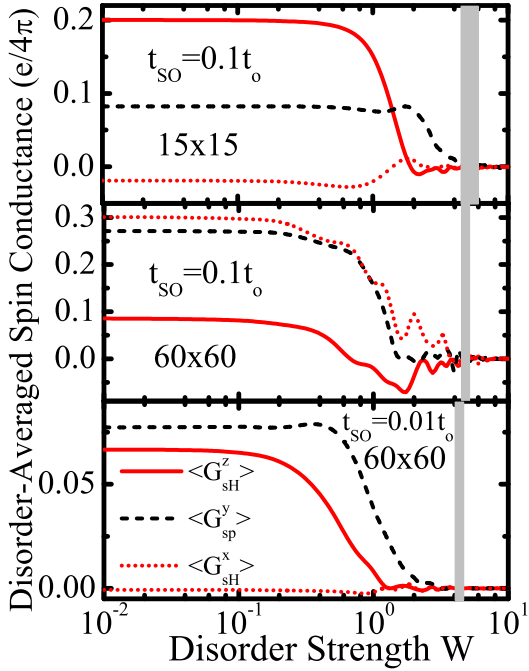,width=2.8in,angle=0}}
\caption{(Color online) The effect of the spin-independent static disorder on the spin Hall  $\langle G_{sH}^z \rangle$, $\langle G_{sH}^x \rangle$ and spin polarization $\langle G_{sp}^y \rangle$ conductances in the four-probe mesoscopic bridges with different strengths of the Rashba coupling $t_{\rm SO}$. In the weak disorder limit, semiclassical mean free path is $\ell \approx 21.5 a t_{\rm o}^2/W^2$, while in the shaded range of  $W$ both the disorder-averaged $\langle G \rangle$ and the typical $e^{\langle \ln G \rangle}$ two-probe charge conductance of the SO coupled 2DEG (attached to two leads) diminish below $\simeq 0.1 \, (2e^2/h)$ due to {\em strong} localization effects.}\label{fig:gh_disorder} 
\end{figure}

Here we shed light on the effect of disorder on the spin Hall current $I_2^s$, whose  
maximum value is set by the SO coupling effects in ballistic transport regime through 
multiprobe mesoscopic structures, by introducing spin-independent static random 
potential $\varepsilon_{\bf m} \in [-W/2,W/2]$  into the 2DEG region of Fig.~\ref{fig:setup} 
and studying the disorder-averaged conductances $\langle G_{sH}^x \rangle$, $\langle G_{sp}^y \rangle$, $\langle G_{sH}^z \rangle$ in the crossover from the {\em quasiballistic} to the {\em localized} transport regime. The inclusion of all transport regimes is made possible by employing the exact single-particle spin-dependent Green function Eq.~(\ref{eq:green}) which encompasses all quantum-interference effects at arbitrary $W$ and $t_{\rm SO}$ in a finite-size device, rather than treating only the lowest order (semiclassical) effects of the disorder~\cite{loss,inoue,shytov,sugimoto} or weak localization quantum corrections.~\cite{chalaev}  Figure~\ref{fig:gh_disorder} suggests that mesoscopic spin Hall conductances are  unaffected by weak disorder, and they  will gradually diminish toward negligible values only at $\alpha k_F \hbar/\tau = (t_{\rm SO} \ell)/(t_{\rm o}a) \simeq 0.1$ which is deep inside the diffusive metallic $\ell < L \ll \xi$ regime. Here  the semiclassical mean free path $\ell \approx 21.5 a t_{\rm o}^2/W^2$ (at the selected Fermi energy $E_F=-3.8t_{\rm o}$) is applicable within the Boltzmann transport regime~\cite{allen} $\ell > a$.

The often quoted mantra---all quantum states of disordered
non-interacting electrons in two-dimensions are localized---means
in practice that the localization length $\xi < \infty$ is finite at
arbitrary disorder strength.~\cite{asada} Thus, the  two-probe (charge)
conductance of a sufficiently large $L \gg \xi$ 2DEG with impurities
will decay exponentially fast $G \sim e^{-L/\xi}$. The two
exceptions are quantum Hall 2DEG (where delocalized states exist
in the center of a Landau level~\cite{meso_hall}) and 2D systems
where strong enough  SO coupling can induce the metallic phase $\xi
\rightarrow \infty$ at weak disorder.~\cite{asada} We delineate
in Fig.~\ref{fig:gh_disorder} the boundaries~\cite{asada} of the 
localization-delocalization transition to demonstrate that spin 
Hall conductances will vanish upon increasing disorder before 2DEG 
is pushed into the realm of strong localization where $G_{sH} \rightarrow 0$ 
is trivially expected.

\section{Mesoscopic spin Hall conductance vs. intrinsic spin Hall conductivity}\label{sec:vs}

Both the intrinsic and the mesoscopic spin Hall effect originate from the SO coupling terms in 
Hamiltonians of clean semiconductor systems. Since conductivity and conductance have the 
same unit in two dimensions, one might na\" ively expect  that $G_{sH}^z \equiv \sigma_{sH}$ since it is tempting to connect the total spin Hall current with  the spin current density integrated over the cross section, $I_2^s=j_y^z L$, and use $E_x = (V_1-V_4)/L$ to find that $G_{sH}^z = (j_y^z L)/(E_x L)=\sigma_{sH}$ should not scale with the system size. If, on the other hand, mesoscopic spin Hall current  in Rashba SO coupled devices  is due to the edge spin currents near the contacts with the longitudinal leads,~\cite{shytov} the same arguments would lead to $G_{sH}^z \propto L_{\rm SO}/L$ which decreases with the 2DEG size. 

In contrast to these  na\" ive expectations, Figure~\ref{fig:gh_rashba} reveals more complicated 
scaling behavior of $G_{sH}^z(L)$ in the  ``mesoscopic'' regime where spin Hall conductance oscillates 
for $L \lesssim L_{\rm SO}$, and reaches asymptotic value $G_{sH}^z \simeq 0.1 e/4\pi$ (up to small  
oscillations around it due to phase-coherent and  ballistic nature of transport) in the ``macroscopic'' 
regime $L \gg L_{\rm SO}$. The semiclassical picture of the deflection of spin densities involving the 
SO ``force'' Eq.~(\ref{eq:so_force}) offers an explanation of these oscillations of the spin Hall 
conductance as being due to the change in sign of such ``force'' deflecting the spin which at the same time is precessing, as elaborated in Sec.~\ref{sec:scaling}. The decay of the SO ``force'' magnitude~\cite{so_force} while  the spin is moving along the wire, which arises due to spin decoherence in ballistic SO coupled systems,~\cite{chao,purity} is responsible for the saturation of spin Hall conductance in large samples with strong SO coupling. 

A closer look reveals further fundamental differences between $G_{sH}^z$ and  $\sigma_{sH}$. The 
intrinsic spin Hall effect,~\cite{murakami,sinova} which is driven by  an external electric field 
penetrating infinite homogeneous systems in the clean limit,~\cite{loss,sugimoto} is essentially a semiclassical phenomenon where spin current is generated by the anomalous velocity (due to the Berry 
phase in  momentum space) of Bloch wave packets without requiring the shift of the electron 
distribution function from equilibrium.~\cite{zhang} Such unusual properties of $j_y^z$ carried by the 
whole SO coupled Fermi sea, which depends  only on the equilibrium distribution function and spin-split 
band structure,~\cite{murakami,sinova} have lead to arguments that the intrinsic spin Hall current is 
an equilibrium current which does not actually transport spin between two points in space.~\cite{zhang,rashba_eq} In fact, such $j_y^z \neq 0$, which does not induce spin accumulation 
or can be employed for spin injection, is found in SO coupled systems without 
any applied electric field~\cite{rashba_eq} or obvious sources of dissipation,~\cite{insulator} which is 
compatible with time-reversal invariance since $j_y^z$ does not change sign under the time-inversion $t \rightarrow -t$ transformation.

On the other hand, the mesoscopic spin Hall current $I_2^s$ is a Fermi-surface quantity at zero 
temperature $T \rightarrow 0$ [i.e., the contribution to $G_{sH}$ in Eq.~(\ref{eq:gh_explicit}) 
from Green functions Eq.~(\ref{eq:green}) evaluated at energies $E<E_F$ is  zero] and a genuine 
nonequilibrium response because no total spin  currents can flow throughout the leads of a 
multiterminal device in equilibrium~\cite{spin_hall_ring,kiselev_theorem} $V_p = {\rm const.}$ (as 
discussed in Sec.~\ref{sec:multiprobe}). Moreover, the induction of $I_2^s$ can never be ``dissipationless''~\cite{murakami,insulator} since  $G_{sH}$ in Eq.~(\ref{eq:gh_explicit}) 
is expressed in terms of the spin-resolved charge conductance coefficients whose non-zero 
values, even in perfectly clean systems where $G_{pq}^{\sigma \sigma^\prime}>0$ is set 
entirely by the sample geometry~\cite{baranger} and interfaces at which SO coupling changes abruptly,~\cite{governale,purity} encode the information about dissipation occuring in remote huge 
reservoirs thermalizing electrons to ensure the steady-state transport. That is, in the Landauer setup  current is limited by  quantum transmission through a potential profile while power is dissipated non-locally in the reservoirs.~\cite{baranger} Note also that in the four-terminal devices of Fig.~\ref{fig:setup}, the 
external bias voltage only shifts the relative chemical potentials of the reservoirs into which 
the longitudinal leads eventually terminate so that electrons do not feel any electric field in 
the course of ballistic propagation through clean 2DEG central region.

\begin{figure}
\centerline{\psfig{file=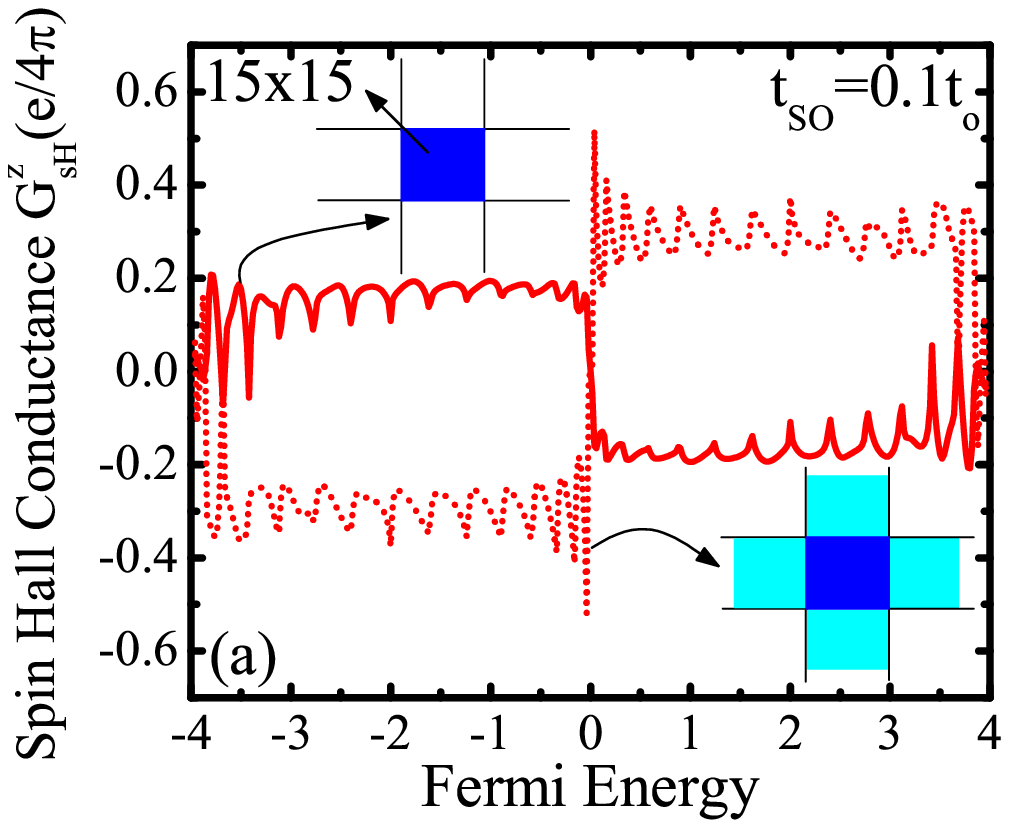,scale=0.7,angle=0} }
\centerline{\psfig{file=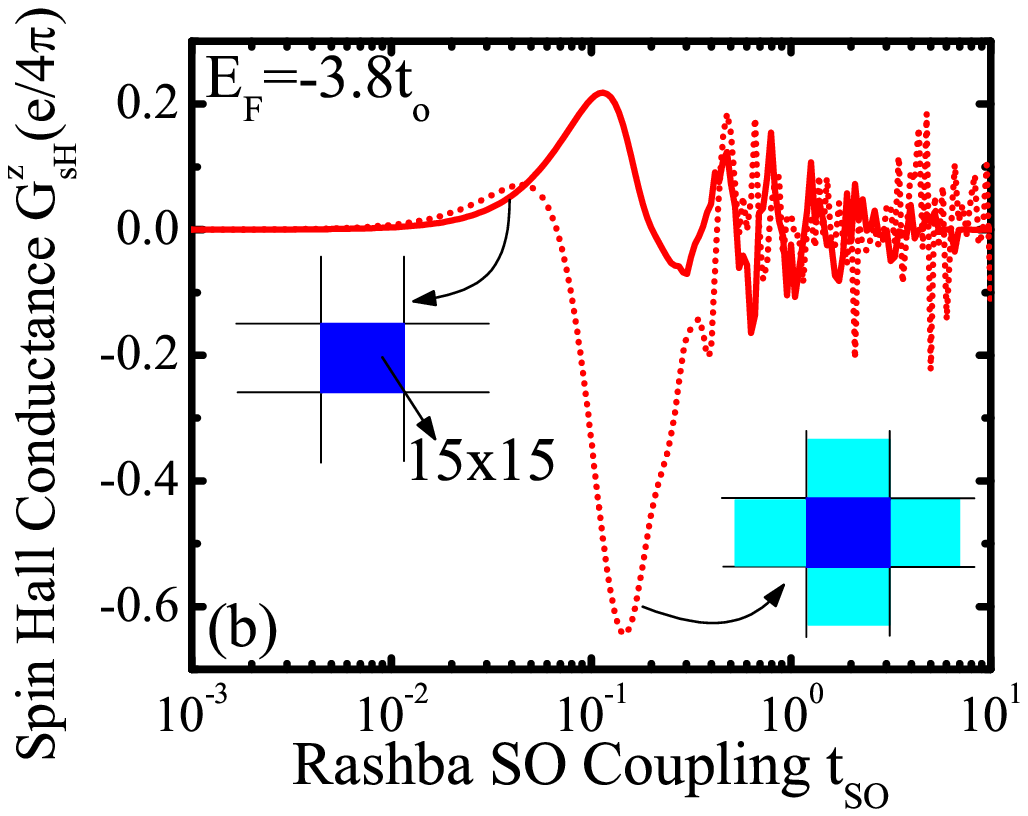,scale=0.71,angle=0}}
\caption{(Color online) The spin Hall conductance $G_{sH}^z$ of a 2DEG of size $15a \times 15a$ with the Rashba SO coupling (dark zone), which is attached to four ideal leads with no Rashba coupling (solid line) 
or four leads where SO coupling is switched on adiabatically via a linear function within a finite 
region (light zone) of length $15a$ of the leads (dotted line).  Panel (a) compares the Fermi energy dependence of  $G_{sH}^z$ for these two different measuring setups, while panel (b) compares their SO 
coupling dependence.} \label{fig:gh_adiabatic}
\end{figure}

The inability to connect bulk conductivity (which relates local current density to the electric field, ${\bf j}=\sigma {\bf E}$) to conductance measured in experiments (which relates the total current to the voltage drop, $I=GV$) is encountered in some charge transport situations as well: (i) in the ballistic regime $\ell \gg L$ conductivity $\sigma$ does not exist as a local quantity and only the conductance $G$ plays a role; (ii) in phase-coherent diffusive conductors quantum corrections to the conductivity emerge which are non-local on the dephasing scale $L_\phi$  that can be much greater than the mean free path (typically  $L_\phi \lesssim 1\mu$m below which the inelastic  processes become suppressed at low enough temperatures $T \ll 1$K), so that concept of local conductivity loses its meaning. Although our devices in Sec.~\ref{sec:ballistic} are both in the ballistic $\ell \gg  L$ and phase-coherent $L < L_\phi$ regime, the principal obstacle in connecting the intrinsic spin Hall conductivity $\sigma_{sH}$ and mesoscopic spin Hall conductance $G_{sH}^z$ lies in the fact that spin is not conserved in SO coupled systems. Thus, the plausibly defined spin current-density operator $\hat{j}_k^i=\frac{\hbar}{4} (\hat{\sigma}_i \hat{v}_k + \hat{v}_k \hat{\sigma}_i )$ (symmetrized product of the Pauli matrices and the velocity operator which yields a Hermitian operator), employed in different computational schemes for the bulk  spin Hall conductivity,~\cite{murakami,sinova,loss,rashba_eq,inoue,chalaev,shytov,sugimoto} lacks 
rigorous theoretical justification and standard physical interpretation  because it does not satisfy the continuity equation with the spin density operator.~\cite{murakami,shytov,rashba_eq} 
 
 While the relation of the spin current density $j_y^z$ (as the  expectation value of the corresponding ``controversial'' spin current-density operator) flowing through the SO coupled system to real spin 
transport and spin accumulation is far from obvious, the pure spin current $\frac{\hbar}{2e}(I_2^\uparrow - I_2^\downarrow)$, which we define within the asymptotic region of the leads with $\alpha \equiv 0$, is conserved quantity so that $I_2^s = {\rm const.}$ does not change on different transverse cross sections throughout the lead 2.  Such pure spin currents flowing through the region with no SO coupling have transparent physical interpretation: If all spin-$\uparrow$ electrons move in one  direction, while an equal number of spin-$\downarrow$ move in the opposite direction, the net charge  current vanishes while spin current can be non-zero. In fact, they have been created and detected in recent optical pump-probe experiments.~\cite{stevens2003}

Even in the semiclassical transport regimes at higher temperatures (where $L > L_\phi$), the presence of SO coupling emphasizes  the demand to treat the {\em whole} device geometry when  studying  the dynamics of transported spin densities.~\cite{wees} For  example, the decay of nonequilibrium spin polarizaions in ballistic or disordered quantum wires is highly dependent on the transverse confinement effects~\cite{purity} or chaotic vs. regular boundaries of clean quantum dots.~\cite{chao} This is to be  contrasted with the 
conventional D'yakonov-Perel' spin relaxation mechanism~\cite{spintronics} in unbounded diffusive systems where the  decay of spin polarization is determined solely by the SO coupling and elastic  spin-independent scattering of charges on the impurities in the bulk.~\cite{chao} Also, the eigenstates~\cite{governale} of SO coupled wires substantially differ from the ones of the infinite 2DEG since ${\bf B}_R({\bf p})$ is almost parallel to the transverse direction~\cite{governale,purity,so_force} (in contrast to the infinite 2DEG where no unique spin quantization axis exists~\cite{sinova}). These transverse confinement effects are found in Sec.~\ref{sec:scaling} to be responsible for the non-zero $[I_2^s]^x \neq 0$ component of the mesoscopic 
spin Hall current, which is quite different from  $j_y^x \equiv 0$ property  of the intrinsic spin Hall effect in infinite homogeneous systems. 

The mesoscopic transport techniques, developed to treat the whole measuring geometry as demanded by 
quantum coherence effects and non-local nature of transport measurements in phase-coherent devices,~\cite{baranger} are well-suited to handle  all relevant details of the spin Hall 
bridges, as applied in Sec.~\ref{sec:multiprobe} to spin transport. We investigate this issues further in 
Fig.~\ref{fig:gh_adiabatic} by studying the effect of measuring geometry on the maximum value of the spin 
Hall  conductance $G_{sH}^z$ in the  four-probe bridges with $L_{\rm SO} \times L_{\rm SO}$ 2DEG, where we employ the leads containing finite region within which the Rashba SO coupling is switched on adiabatically (via linear function) from $t_{\rm SO}=0$ to the value it acquires in the 2DEG. Compared with our standard setup from Fig.~(\ref{fig:setup}), the usage of such electrodes would enhance the spin Hall effect since reflection at interfaces where $t_{\rm SO}$ changes abruptly is greatly reduced.

\section{Concluding Remarks}\label{sec:conclusion}

In conclusion, we have delineated features of a novel type of spin Hall effect in four-terminal 
mesoscopic structures where unpolarized charge current driven through the longitudinal ideal 
(with no spin and charge interaction) leads  attached to clean semiconductor region  with 
strong enough homogeneous SO coupling induces  a pure spin current in the transverse voltage 
probes (with zero net charge flow through them). The spin carried by the transverse  spin Hall current 
in devices where electrons are subjected to the Rashba-type of SO coupling within the central region 
has  both out-of-plane and in-plane non-zero components of its polarization vector.  The spin Hall current depends on the strength of the Rashba SO coupling. Furthermore, the maximum value of the 
spin Hall conductance $G_{sH}^z$ is always achieved for the square samples of the size $L_{\rm SO} \times L_{\rm SO}$, where spin precession length $L_{\rm SO}$ can be tuned by changing the Rashba SO coupling via 
the gate electrode covering the 2DEG.

Although apparently similar to recently predicted intrinsic spin Hall effect (as a semiclassical phenomenon  
in infinite homogeneous clean SO coupled systems), the mesoscopic pure spin Hall current predicted here has fundamentally different properties: It is a genuine nonequilibrium and  Fermi-surface quantity which depends on the whole device (i.e., interfaces and boundaries) and measuring geometry (e.g., it can be enhanced by the leads where SO coupling is adiabatically switched on within a finite region). Its non-trivial finite-size scaling regimes in  samples smaller and larger than $L_{\rm SO}$ highlight the importance of processes~\cite{so_force}  involving spin dynamics on this mesoscale for the generation of spin Hall current in ballistic multiprobe nanostructures. In contrast to the bulk intrinsic spin Hall conductivity of the Rashba spin-split 2DEG, mesoscopic spin Hall conductances are able to survive weak scattering off impurities, gradually decaying within the metallic diffusive transport regime. 

{\em Note added.---} Upon completion of this work we have become aware of Ref.~\onlinecite{sheng}
where similar mesoscopic approach to spin Hall effect has been undertaken for different four-probe 
structure---an infinite Rashba SO coupled wire with two transverse ideal leads attached---and 
analogous conclusions have been reached regarding its resilience to disorder effects.

\begin{acknowledgments}
We are grateful to J. Inoue and A. Shytov for important insights. This research was supported in part by ACS grant No. PRF-41331-G10.
\end{acknowledgments}


\end{document}